# Determination of the electrostatic lever arm of carbon nanotube field effect transistors using Kelvin Force Microscopy

David Brunel[1], Dominique Deresmes[1], and Thierry Mélin[1,a]

[1] *Institute of Electronics, Microelectronics and Nanotechnology, CNRS-UMR 8520, ISEN Department, Avenue Poincaré, BP 60069, 59652 Villeneuve d'Ascq Cedex, France*

[a] Electronic mail: thierry.melin@isen.iemn.univ-lille1.fr

Abstract

We use Kelvin Force Microscopy (KFM) to study the electrostatic properties of single-walled Carbon Nanotube Field Effect Transistor devices (CNTFETs) with backgate geometry at room temperature. We show that KFM maps recorded as a function of the device backgate polarization enable a complete phenomenological determination of the averaging effects associated with the KFM probe side capacitances, and thus, to obtain KFM measurements with quantitative character. The value of the electrostatic lever arm of the CNTFET is determined from KFM measurements, and found in agreement with transport measurements based on Coulomb blockade.

*PACS*: 07.79.-v, 68.37.Ps, 81.07.De, 85.35.Kt

Because of their electronic properties [1], carbon nanotubes have been widely used as field effect transistors (CNTFETs) since the last decade [2-7]. So far however, only a few studies have been performed in order to characterize the local electronic properties of operating nanotube devices. One reason for this is that these investigations can only be achieved using electrical techniques derived from atomic force microscopy such as Electrostatic Force Microscopy (EFM) or Kelvin Force Microscopy (KFM), which is still a field under development. Significant information can however be extracted from such investigations with respect to devices, as seen from available studies on *e.g.*: charge transfers at metal-nanotube interfaces in air or vacuum [8], transport regimes in connected nanotubes [9], the assessment of contact resistances [10], or hysteresis effects in CNTFETs [11].

A central issue for such studies is to extract *quantitative* information from the measurement of surface potentials, because the microscope cantilever tips actually "feel" the electrostatic properties of the device through a series of parallel capacitances (see Figure 1 for illustration). In KFM, the primary effect of the side capacitances is to degrade the lateral resolution, but also, as pointed out by Jacobs *et al.* [12], to prevent from direct quantitative local surface potential measurements due to the averaging of the local surface potentials associated with each of the parallel capacitances [13].

In this Letter, we use KFM to study the electrostatic properties of CNTFETs, and show that KFM maps recorded as a function of the device backgate polarization enable a complete determination of the averaging effects associated with the KFM probe side capacitances. This quantitative analysis – which might be easily generalized to other types of nanodevices – is used here to measure the electrostatic lever arm of the CNTFET, which is found in agreement with transport studies based on low-temperature Coulomb blockade measurements [14,15].

CNTFETs have been fabricated using commercial single walled carbon nanotubes (Nanocyl, Belgium) dispersed in dichloromethane, and randomly deposited on a 320 nm thick silicon dioxide layer thermally grown on a highly *p*-doped silicon wafer. Electron beam lithography has been used to design the 20 nm thick Palladium source and drain contacts (see the topography image of the CNTFET in Figure 1). Transport and KFM measurements have been carried out at room temperature in a Veeco Dimension 3100/Nanoscope IV microscope placed under dry nitrogen atmosphere. We used $Pt_{0.95}/Ir_{0.05}$ metal-plated cantilevers with spring constant ~3 N/m and a resonance frequency ~70 kHz. Topography and KFM data have been mapped using a standard two-pass procedure, in which each topography line acquired in tapping mode is followed by the acquisition of KFM data in a lift mode, with the tip scanned at a distance z~80 nm above the sample so as to discard short-range surface forces and be only sensitive to electrostatic forces.

In the KFM mode, a $dc + ac$ bias $V_{dc} + V_{ac}\sin(\omega t)$ is applied to the cantilever (here $V_{ac} = 2\ V$), with $\omega/2\pi$ at the cantilever resonance frequency. This excitation generates a capacitive force component $F_\omega$ (and thus a cantilever oscillation) at the angular frequency $\omega$, of amplitude $F_\omega = \partial C / \partial z (V_{dc} - V_S) V_{ac} \sin(\omega t)$. In this expression, $C(z)$ is the tip-substrate capacitance, and $V_S$ denotes the local surface potential, which accounts for the work function difference between the tip and substrate, the presence of local charges in the tip-substrate capacitance, and the local electrostatic potential over the device. Experimentally, $V_S$ is measured using a feedback loop which sets to zero the cantilever oscillation amplitude at $1\omega$ (and thus $F_\omega$) by adjusting the tip dc bias $V_{DC}$. This potential (*i.e.* the output of the KFM loop) is here after denoted $V_{KFM}$. It is simply equal to $V_S$, in absence of side capacitances.

To describe the effect of side-capacitances, we follow the approach of Ref. [12] and limit ourselves to the following capacitances (see Figure 1) when the tip is scanned over the nanotube: $C_{NT}$ is the KFM tip/nanotube capacitance associated with the nanotube potential $V_{NT}$; $C_G$ is the tip/backgate side-capacitance associated with the backgate bias $V_G$; finally, $C_D$ and $C_S$ are the capacitive couplings between the KFM tip and the drain and source contact pads at the potential $V_D$ and $V_S$, respectively, which occur between the pyramid part of the cantilever tip and the metallic contacts on the surface, and thus, at a few $\mu m^2$ scale.

In presence of the side capacitances, and taking $C = C_{NT} + C_G + C_D + C_S$, the $V_{KFM}$ voltage can be written as a weighted average of the surface potentials $V_{NT}$, $V_G$, $V_D$ and $V_S$ [12]:

$$V = \alpha_{NT} V_{NT} + \alpha_G V_G + \alpha_D V_D + \alpha_S V_S \qquad (1)$$

in which $\alpha_{NT} = (\partial C_{NT}/\partial z)/(\partial C/\partial z)$, $\alpha_G = (\partial C_G/\partial z)/(\partial C/\partial z)$, $\alpha_D = (\partial C_D/\partial z)/(\partial C/\partial z)$ and $\alpha_S = (\partial C_S/\partial z)/(\partial C/\partial z)$ are the normalized capacitance gradients associated with $C_{NT}$, $C_G$, $C_D$ and $C_S$, respectively. In addition, the action of a backgate voltage shift $\Delta V_G$ on the nanotube during the CNTFET operation is expressed by $\Delta V_{NT} = \beta \cdot \Delta V_G$, in which $\beta$ is the device electrostatic lever arm. The aim will be to determine the values of the $\alpha$ coefficients phenomenologically (*i.e.* without determining $C_{NT}$, $C_G$, $C_D$ and $C_S$), and of the lever arm $\beta$. In practice, the values of the weights $\alpha$ may also depend on the position along the nanotube [13]. They will be determined here at the middle of the nanotube. A first relation comes from the normalization of the $\alpha$ coefficients: $\alpha_{NT} + \alpha_G + \alpha_D + \alpha_S = 1$. Four independent equations therefore still need to be established from experiments to solve for the coefficients $\alpha$ and $\beta$.

To do so, we start from the KFM images of a polarized nanotube device (see Figure 2), in which one monitors the changes in the KFM image of a CNTFET when biased using

$V_{DS} = -3\ V$ *i.e.* in a diffusive transport regime (see Figure 2b and c). The sections of the KFM potential along the nanotube are shown in Figure 2d, together with the difference of the surface potential for $V_{DS} = -3\ V$ and $V_{DS} = 0\ V$ shown in Figure 2d. We first monitor the change in the KFM potential at the source upon the drain bias, which is found less than 30 mV, indicating that $\alpha_D \sim 1\%$ at the source. This value will be kept constant in the following. We then extract from Figure 2d the voltage drop $\Delta V = 600 \pm 50\ mV$ along the device, using a linear fit so as to circumvent the issue of apparent voltage drops at contacts (see Figure 2d), due to the local change in the $\alpha$ coefficients at the contacts, and/or to the effect of contact resistances. From Eq. (1), the ratio between the nanotube voltage drop $\Delta V$ measured from KFM and $V_{DS}$ directly equals $\alpha_{NT} + \alpha_D$, which gives $\alpha_{NT} = 0.19 \pm 0.02$.

The second step consists in imaging the CNTFET as a function of the gate polarization $V_G$. The corresponding KFM images are presented in Figure 3 for backgate voltages between $V_{GS} = -3\ V$ and $V_{GS} = -12\ V$, corresponding to the activation regime of the *p*-type CNTFET, as seen from its transfer characteristics shown in the inset of Figure 4. We show in Figure 4 a plot of the KFM surface potentials $V_{KFM}$ measured *(i)* above the nanotube and *(ii)* above the SiO$_2$ layer outside the nanotube (the positions of the measurement are indicated by a circle and a triangle in the topography image of Figure 3a). In both situations, the evolution of the KFM surface potential is linear with $V_G$ with a sub-unity slope, and furthermore exhibits shows a slower slope ($0.75 \pm 0.02$) when the measurements are taken above the nanotube, as compared to the SiO$_2$ surface ($0.89 \pm 0.02$). This is consistent with Eq. (1) giving a slope $\alpha_{NT}\beta + \alpha_G$ (with $\beta < 1$) over the nanotube, because the nanotube potential $V_{NT}$ varies as $\beta \cdot V_G$. When the KFM measurement is taken on the oxide outside the nanotube, the nanotube potential $V_{NT}$ has to be replaced by the backgate potential $V_G$, so that the slope of the KFM surface potential as a function of $V_G$ becomes $\alpha_{NT} + \alpha_G$.

Starting from $\alpha_{NT} + \alpha_G = 0.89$, one obtains $\alpha_G = 0.70$. Using the normalization relation, one finds $\alpha_D + \alpha_S = 0.11$, and thus $\alpha_S \sim 0.10$. All the $\alpha$ coefficients are measured here with a typical accuracy of $\pm 0.02$. The larger value of $\alpha_S$ compared to $\alpha_D$ is explained by the presence of an additional electrode connecting the nanotube close to its source side, though not visible in the KFM image. The lever arm $\beta$ beta can then be determined using the relation $\alpha_{NT}\beta + \alpha_G = 0.75 \pm 0.02$, giving finally $\beta = 0.26 \pm 0.15$. The relatively large error bar for $\beta$ comes from the small weight of $\alpha_{NT}\beta$ in the above relation. It could be improved for instance by using smaller tip-substrate separations (*e.g.* using non-contact atomic force microscopy in vacuum), so as to increase the value of $\alpha_{NT}$ in the weighted average of the side capacitance derivatives. To confirm the value of the lever arm found from KFM experiments, we subjected the carbon nanotube of a CNTFET device (not shown here) to a charge injection experiment (see Ref. [16] for details of the procedure), known to generate a homogeneous linear charge in the SiO$_2$ layer along the nanotube, with density in the $10-100\,e/\mu$m range [16]. We observed both a rigid shift $\Delta V_G$ of the CNTFET transfer characteristics after the charge injection, together with a $\Delta V_{KFM}$ shift of the carbon nanotube surface potential after charging, with a ratio $|\Delta V_{KFM}/\Delta V_{GS}| = 0.06 \pm 0.01$. In the side-capacitance model used in our work, this ratio equals $\alpha_{NT}\beta$ as seen from Eq. (1). The above determined values of $\alpha_{NT} = 0.19 \pm 0.02$, and $\beta = 0.26 \pm 0.15$ correspond to $\alpha_{NT}\beta = 0.05 \pm 0.03$, and are thus in agreement with the $|\Delta V_{KFM}/\Delta V_{GS}|$ ratio found from charge injection experiments. This value of the CNTFET electrostatic lever arm is also in agreement with measurements obtained from Coulomb oscillations in the low temperature transport in µm-long nanotube devices

(Radosavljević *et al.*[14], Babić *et al.*[15]), showing $\beta$ values in the $0.2-0.4$ range. Further developments should enable to measure the electrostatic lever arm directly at the nanotube contacts, which governs the carrier injection into the CNTFET.

In conclusion, we have studied in this work the electrostatic properties of CNTFETs using Kelvin force microscopy. We showed the possibility to achieve quantitative measurements through the experimental determination of the averaging effects associated with the KFM probe side capacitances, allowing the determination of the CNTFET electrostatic lever arm. This work was done in the framework of the CNRS GDR-E No.2756, and supported in part by the ANR Grants No. JC05_46152, 06-NANO-070, and PHC project No.17806NL. The authors would like to acknowledge valuable discussions with M. Zdrojek, and the technical staff involved in the device processing.

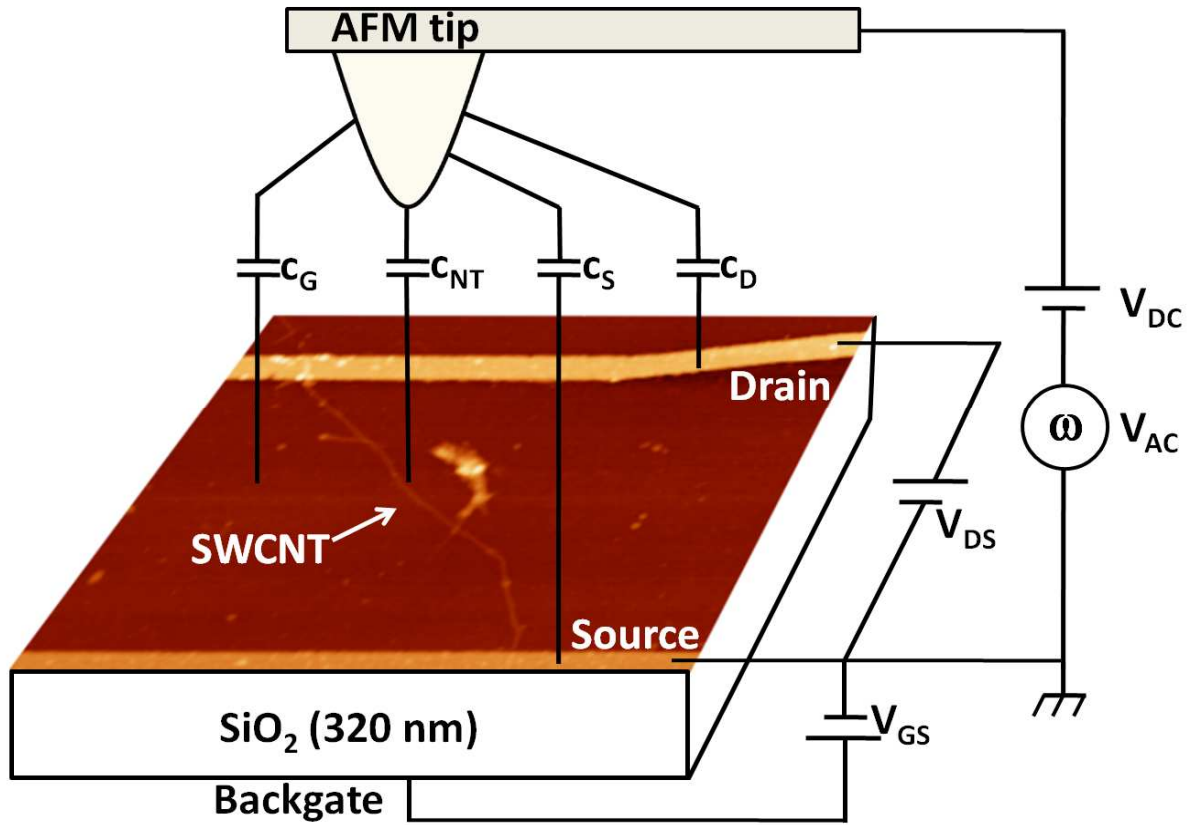

*Figure 1: (Color online) Schematics of the KFM detection set-up, showing the $5\mu m \times 5\mu m$ atomic force microscopy topography image of a CNTFET in backgate geometry. The tip is lifted with respect to the surface for sake of clarity (actual tip-substrate distance separation ~80 nm). Source, drain, and gate potentials are denoted $V_S$, $V_D$, and $V_G$, respectively. The cantilever bias used for the KFM measurement is $V_{dc} + V_{ac}\sin(\omega t)$. The tip/nanotube $C_{NT}$ and side capacitances $C_G$ and $C_D$ and $C_S$ are represented schematically (see the text for a full description).*

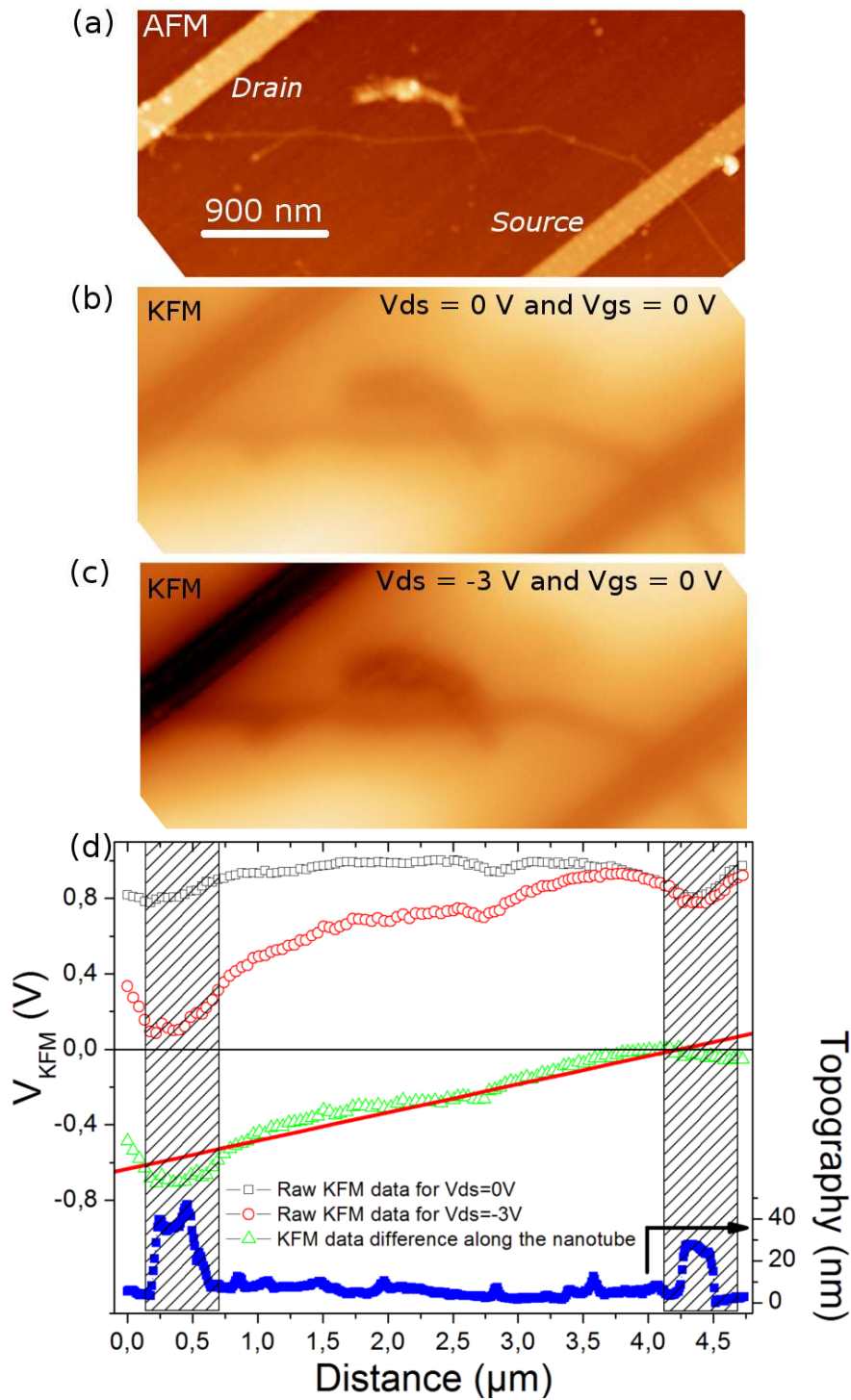

*Figure 2: (Color online) (a) AFM image of the CNTFET of Figure 1. The scale bar is 900 nm. (b) Associated KFM image (tip-substrate distance ~80 nm, $V_{ac} = 2\ V$) with $V_{DS} = 0V$ and $V_{GS} = 0V$. The color scale is 1.5 V. (c) Similar image, with $V_{DS} = -3\ V$ and $V_{GS} = 0V$. (d) Sections of the KFM signals of (b) and (c) taken along the carbon nanotube, with $V_{DS} = 0V$ (black circles) and with $V_{DS} = -3\ V$ (red squares). The blue line corresponds to the associated topography section showing the CNTFET source and drain contacts.*

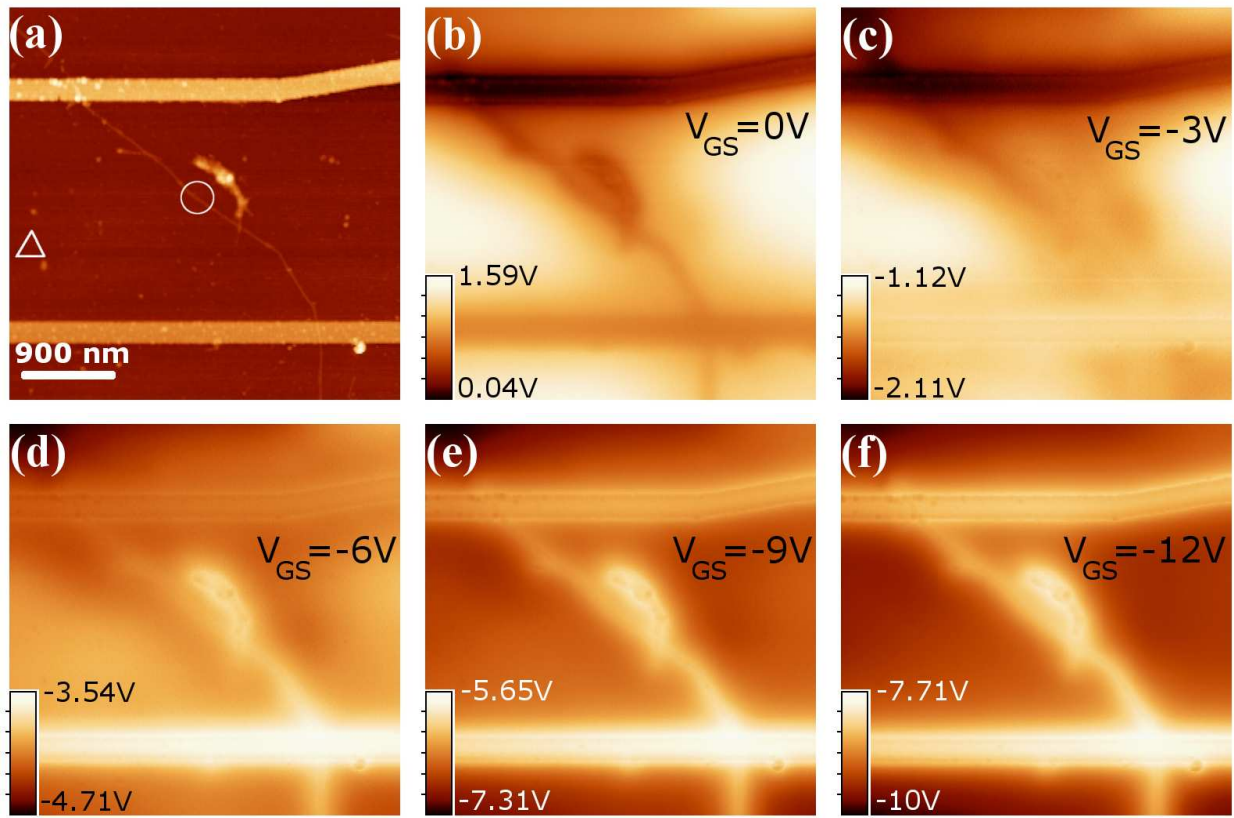

*Figure 3: (Color online) (a) Topography image of the CNTFET (scale bar: 900 nm). (b-f) Associated KFM images recorded for $V_{DS} = -3\ V$ and with (b) $V_{GS} = 0V$ (c) $V_{GS} = -3\ V$ (d) $V_{GS} = -6\ V$ (e) $V_{GS} = -9\ V$ and (f) $V_{GS} = -12\ V$. The color scales are shifted for sake of clarity.*

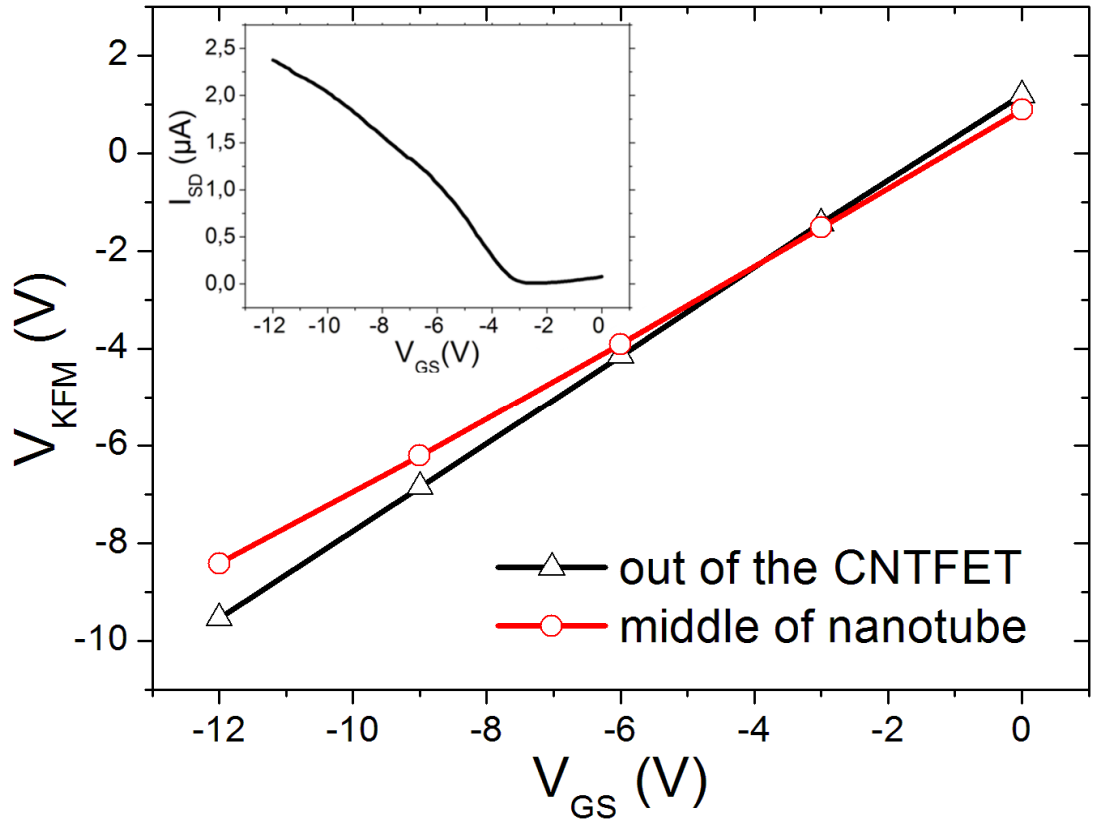

*Figure 4: (Color online) Inset: CNTFET transfer characteristics. Main figure: plot of KFM signal $V_{KFM}$ for $V_{DS} = -3\,V$, as a function of the gate potential $V_{GS}$. Data have been recorded on the nanotube (circles) and out of the nanotube (triangles). The position of the measurements is indicated in the topography image in Fig. 3a.*